# Reasonableness discussion and analysis for Hyperledger Fabric configuration

Song Hua
*Information Technology Laboratory*
*Fujitsu Research & Development Center*
Suzhou, China
huasong@cn.fujitsu.com

Shenbin Zhang
*Information Technology Laboratory*
*Fujitsu Research & Development Center*
Suzhou, China
zhangshenbin@cn.fujitsu.com

Bingfeng Pi
*Information Technology Laboratory*
*Fujitsu Research & Development Center*
Suzhou, China
winter.pi@cn.fujitsu.com

Jun Sun
*Information Technology Laboratory*
*Fujitsu Research & Development Center*
Suzhou, China
sunjun@cn.fujitsu.com

Kazuhiro Yamashita
*Software Laboratory*
*Fujitsu Laboratories*
Kawasaki, Japan
y-kazuhiro@jp.fujitsu.com

Yoshihide Nomura
*Software Laboratory*
*Fujitsu Laboratories*
Kawasaki, Japan
y.nomura@jp.fujitsu.com

*Abstract*—Blockchain, as a distributed ledger technology, becomes more and more popular in both industry and academia. Each peer in blockchain system maintains a copy of ledger and makes sure of data consistency through consensus protocol. Blockchain system can provide many benefits such as immutability, transparency and security. Hyperledger Fabric is permissioned blockchain platform hosted by Linux foundation. Fabric has various components such as peer, ordering service, chaincode and state database. The structure of Fabric network is very complicated to provide reliable permissioned blockchain service. Generally, developers must deal with hundreds of parameters to configure a network. That will cause many reasonableness problems in configurations. In this paper, we focus on how to detect reasonableness problems in Fabric configurations. Firstly, we discuss and provide a reasonableness problem knowledge database based on the perspectives of functionality, security and performance. Secondly, we implemented a detect tool for reasonableness check to Fabric. Finally, we collect 108 sample networks as the testing dataset in the experiment. The result shows our tool can help developers to locate reasonableness problems and understand their network better.

*Keywords—Permissioned blockchain, network configuration, Hyperledger Fabric*

## I. Introduction

In the past few years, blockchain becomes one of the most popular technologies in the world. As the success of Bitcoin provided by Nakamoto[1], many organizations and companies increase their interest in blockchain. A blockchain network can be defined as an immutable distributed ledger maintained by multiple nodes. Every node stores the whole copy of ledger and makes sure the consistency of all data by using consensus protocol. Each block includes a hash that bind to the preceding block. These characteristics guarantee that the blockchain system is very difficult to be victimized.

Blockchain can be classified into two types: public and permissioned blockchain. A public blockchain, such as Bitcoin and Ethereum, can allow any participant to join the network and access any information in the ledger. PoW[1] or PoS [2] consensus protocol is used by most of the public blockchain platforms. Public blockchain system is very suitable for cryptocurrency applications. But for the common business logic in the enterprise, the access control and membership management are required. A permissioned blockchain is another type of blockchain system. It allows only authenticated or permissioned participants to access ledger data and create new transactions.

Hyperledger Fabric[3] is one of the permissioned blockchain platforms. It's open sourced and developed by Linux foundation. Fabric consists various components such as peer nodes, clients, ordering service, membership, and Chaincode[20]. Each component has different role for different purpose. The transaction flow contains four main phases, endorsement, ordering, validation and committing. All the components need to be customized and configured before network startup. Developers must deal with hundreds of parameters around all components to bootstrap their customized Fabric network. Even worse, many of these parameters are correlated with each other.

In this paper, we focus on reasonableness problems of network configuration in Fabric. The reasonableness problems are imperfections that cannot satisfy users' requirements during network configuration, which may cause low efficiency, insecurity or even functionality missing. Developers need a lot of experience to avoid these problems before starting up the network. Therefore, we provide a solution and a tool to help the developers to detect reasonableness problems in their network configurations.

The rest of the paper is organized as follows: Section II discusses related work. Section III introduces the background of Fabric structure and configuration. Section IV describes reasonableness problems in Fabric configuration. Section V shows our detection tool for reasonableness problem. Section VI is the experiment and result. Finally Section VII gives the conclusion and future work.

## II. RELATED WORK

There some solutions and discussion focused on the optimization of Fabric. For performance optimization, Gorenflo et al.[12] increased transaction throughput of Fabric by re-architecting and modifying the framework and components. Thakkar et al. [13] found that the endorsement policy verification, the sequential policy validation of transactions and the state validation and commit with CouchDB are the three major bottlenecks of Fabric through their elaborated experiments. And they also introduced some simple optimizations such as aggressive caching for endorsement policy verification. Baliga et al.[14] took experimental approach to understand performance characteristics of Fabric. In [15], Sukhwani et al. presented a performance model using Stochastic Reward Nets (SRN) to compute the performance index. Javaid et al.[16] re-architected the validation phase of Fabric based on their analysis of fine-grained latency to increase transaction committing performance.

For functionality and security, Andola et al. [17] discussed the two security limitations of Fabric related with DoS attack and wormhole attack, then provided methods to remove the weakness based on communication verification. Vukolic et al.[18] also discussed some of limitations in Fabric, and presented some re-design advise of Fabric architecture. Yewale et al. [19] mentioned the complexity of Fabric network deployment and created an environment using Kubernetes.

Although these papers are very enlightening, but the solutions for optimizing Fabric all focused on how to restructure or modify Fabric framework. There is no paper that explores or summarize the reasonableness problems from the perspective of network configuration.

## III. BACKGROUND: HYPERLEDGER FABRIC STRUCTURE & CONFIGURATION

Fabric implements complex architecture and multiple different components to provide its high adaptation feature.

### A. Fabric components

- Peer

Peer is a fundamental component of a blockchain network. It is a kind of node that hosts the ledger and smart contracts in Hyperledger Fabric. Peers can be grouped into channels to manage different ledgers individually. In a single channel, each peer can hold a whole copy of the ledger and smart contracts. Fabric network consists of multiple organizations. Peers are owned by these organizations by identity and certifications, which needs to be configured at the beginning of the network setup.

There are two major roles for peer. 1) Endorser. In the beginning of the Fabric transaction flow, applications generate a transaction proposal and send it to each of the required set of peers for endorsement. Every endorsing peer executes the smart contract independently to generate the proposal response. It will not apply real update to ledger, but contains the required signatures of related peers and their independent read/write set from execution of smart contract. 2) Committer. All committers in a channel receive the ordered blocks from the ordering service and then update the specific ledger. Before updating ledger, committers also verify whether every transaction is valid or not based on several rules[3].

- Ordering service and orderer

Orderer nodes, as the ordering service supporters, sort the transactions submitted from applications after endorsement. The orderer nodes receive the data using atomic broadcast protocol[3,7].

The ordering service implements the consensus protocol to order the transactions. It can use Solo, Kafka[4] or Raft[5] as consensus method in Fabric. Solo is suggested to be used for research only. It runs as a process on a single orderer node, and cannot support crash fault tolerant. Kafka is implemented on several nodes outside of the orderer nodes. Raft is a crash fault tolerant ordering service based on the Raft protocol in etcd. The main difference from Kafka is that in Raft, everything is embedded into the orderer nodes.

Fabric provides several configuration parameters such as block timeout and block size in ordering service for customized purpose.

- Membership service provider

Membership Service Provider (MSP) is a Fabric component which manages the identities of all participants in this blockchain network[3]. The identities of participants are implemented by Certificate Authority (CA), Public Key Infrastructure (PKI). MSP abstracts all cryptographic operations such as issuing and validating the certificates. Developers can use MSP to define their required identities related with the organizations, peers, ordering service and users or applications. All the nodes, users and clients use digital certificates to verify each other and communicate with each other.

- Smart contract

Smart contract is one of the key components of Hyperledger Fabric. As a blockchain system, a smart contract defines series of executable logic which are stored in the ledger. Fabric uses a general-purpose programing language based smart contract called chaincode to fulfill the business logic and access the ledger data. There are 2 types of chaincode, a general chaincode provided by developer or user and a system chaincode hosted by Fabric framework.

Fig 1 shows the chaincode invocation flow in Fabric. An invocation of chaincode contains 5 main phases. In phase 1 and 2, applications send transaction proposal to specific set of peers to execute the chaincode for endorsement and get responses. In phase 3, applications send transaction with endorsement results to ordering service to package transactions into blocks. In phase 4 and 5, all peers in channel pull blocks from ordering service, validate and commit all the transactions.

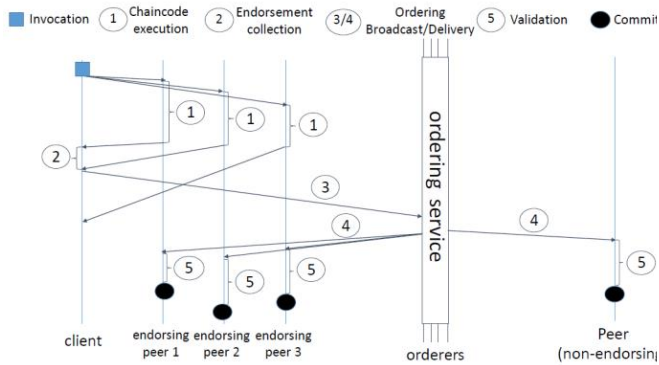

Fig. 1. Transaction flow of chaincode invocation [3]

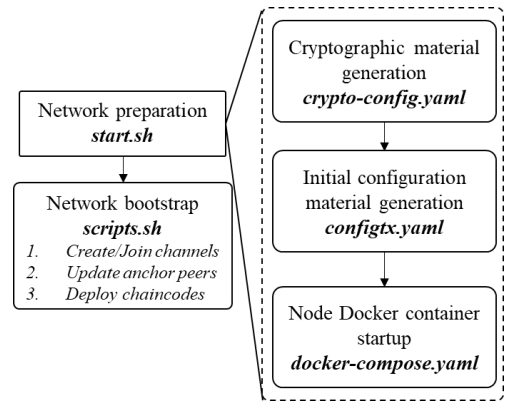

Fig. 2. Fabric bootstrap flow

- State database

A chaincode persists a set of data which called world state that contains all current state values of objects. All world state data are organized as key value pairs. Chaincodes can use put, get and delete operations to interact with world states. The latest values of all keys in chaincodes are stored in a state database.

In Fabric, there are two official state database options, LevelDB and CouchDB. LevelDB is embedded into the peer node process. It stores world state as the key-value pairs. LevelDB is the default state database in the official document of Fabric. CouchDB is another optional state database that can support rich data query function when the chaincode data is constructed as JSON format. Compared to LevelDB, the shortcoming is that the CouchDB has lower efficiency on data processing.

CouchDB runs as a separate process outside the peer process. So there are more things to do in the setup, management and operations. If there are additional complex rich query requirements, developers need to consider the migration to CouchDB from LevelDB. Otherwise developers should keep using LevelDB for high performance.

### B. Fabric configuring, building and bootstrap

In the official documents, there is a recommended solution to build a Fabric network. In general, developer need several *Yaml*[9] files, shell scripts as the configuration and building materials, and use Fabric binary tools and Docker to bootstrap all the necessary components. Docker supports a container environment to Fabric for easier deploy and maintenance[6]. Fig 2 shows the main configuration and scripts structure to build a general network.

- Cryptographic material generation

Fabric provides a tool named *cryptogen*[21] to generate the required cryptographic material, these are x509 certificates and signing keys, for the whole network entities. *Cryptogen* consumes a file, usually named *crypto-config.yaml* to generate a set of certificates and keys for the organizations, peers, orderers and users. The *crypto-config.yaml also* contains the basic topology of the network.

In fabric, any information should be signed by private key and verified by corresponding public key.

- Initial configuration material generation

*Configtxgen* tool is used to generate the following necessary initial configuration materials.

*1) Genesis block*
The genesis block is the first block in the ledger. It's a configuration block that initializes the ordering service and the original network structure.

*2) Channel configuration transaction*
This transaction will be broadcast to the ordering service after network startup for the channel creation operations. It defines and determines channels for this network.

*3) Anchor peer transations*
These transactions will specify anchor peer for each organizations on this channel one by one. Anchor peers are used by gossip to make peers in different organizations know about each other. There must be at least one anchor peer exist in one channel, and it's recommended that there should be set of anchor peers in every organizations for crash fault tolerant and high performance.

*Configtxgen*[22] consumes a file named *configtx.yaml*. *Configtx.yaml* specifies the definitions of the target network. The definitions contains organizations, peers, policies, ACLs, capabilities and other structure configurations.

- Node Docker container startup

By default, Fabric uses Docker-compose to startup component processes in batches. Docker-compose consumes a file named *docker-compose.yaml*. *Docker-compose.yaml* defines the configurations of every Docker containers for thepeer nodes, orderer nodes, CouchDB nodes, CA servers, CLI container and other components.

The above three steps usually can be coded in a shell script named *start.sh*.

- Fabric network structure bootstrap

The last phase of network startup is to bootstrap the Fabric network structure. This phase contains 3 main steps.

1) Create & join channel

In order to create channels, developer put the channel configuration trans mentioned beofre online. Then they can choose a set of peers to join in channels, according to the requirements.

2) Update anchor peers

Anchor peers must be specified separately in each channel. The method of updating anchor peers is to put the anchor peer transaction materials on the ledger. Submitting all the transactions on the ledger can update channel configuration to add new anchor peers.

3) Deploy chaincodes

The third phase is to deploy the chaincodes to specific channels. In Fabric, all applications must interact with blockchain through chaincode. Developers need to install chaincodes on every peer which will endorse the invocations. Also, developers must decide and set the endorsement policy of every chaincode. Endorsement policy defines the set of organizations that are required to endorse a chaincode invocation transaction. Transactions that are not satisfied endorsement policy will be set to invalid by the committing peers.

These three operations can be put in a shell script named scripts.sh. Different with start.sh, scripts.sh will run in the Fabric CLI container.

TABLE II.    REASONABLENESS CHECK PATTERNS

| Category | Reasonableness problems | Comment |
|---|---|---|
| Functionality | CouchDB vs LevelDB | CouchDB vs LevelDB |
| | Inconsistent parameters | Inconsistent configurations between different sources |
| | Parameter hardcoded | Hardcoded parameters increase cost of debugging and maintenance |
| | Component missing | Configuration integrity |
| | Yaml syntax | - |
| | Docker compose file syntax | - |
| Performance | BlockTime / BlockSize | Configuration of ordering service |
| | Complex chaincode endorsement policy | Too complex, leads to low efficiency |
| Security | Simple chaincode endorsement | Too simple, leads to low security |
| | TLS on/off | TLS off leads to low security in data transportation |
| | State database security | State database authentication information missing |
| | Consensus mechanism | Solo - None CFT Kafka - CFT, hard to governance Raft - CFT |

IV. UNREASONABLE STATEMENT IN FABRIC

Through our experience, we found there are several reasonableness problems. These reasonableness problems can be categorized into 3 types. And a reasonableness knowledge database is generated for these problems. Overall, Table 1 shows all the unreasonable problems. We will discuss them in detail.

A. Functionality

Some of the unreasonable configurations will cause functionality problem for the network. Functionality problems may include absence and failure of function or even crash of the network. The followings are some functionality reasonableness problems which may be caused in Fabric network.

- Adoption of CouchDB vs LevelDB

There are two types of state database for Fabric up to version 1.4.0 – CouchDB and LevelDB.

As we introduced before, the advantages of LevelDB is it has high performance and ease of maintenance. Meanwhile CouchDB can support rich query.

For the developers and administrators, they must choose the suited state database based on their real specific requirements and scenarios. Unreasonable choice of state database will cause low efficiency and limited functions.

- Inconsistent parameters

There are hundreds of parameters in Fabric network configuration. The developers can pre-set and modify any parameters before they construct the whole network based on some configuration files. In these parameters, some are correlated with each other. If there are inconsistency parameters, usually it will cause some problems in network constructing process.

For example, the DOMAIN information must be consistent both in *crypto-config.yaml* file and *docker-compose.yaml*. Fig. 3 shows the instance configurations in a specific Fabric network.

In this two Yaml files, the DOMAIN *org2.example.com* must be consistent otherwise there could be potential problems in the network.

Listing 1: Example of Docker-compose.yaml

```
1.    container_name: company.org.consortium.com
2.    image: hyperledger/fabric-peer:$IMAGE_TAG
3.    environment:
4.      - CORE_PEER_ID=auditor.org.consortium.com
```

Listing 2: Example of Crypto-config.yaml

```
1.    Name: org
2.    Domain: org.consortium.com
3.    Specs:
4.      - Hostname: company
```

Fig. 3 Example of inconsistent parameters

Listing 3: Example of parameters hardcoded

```
1. CA0:
2.   command: sh '** ./3231ea0d_sk'
```

Listing 4: Example of parameters not hardcoded

```
1. CA0:
2.   command: sh '** ./${PRIVATE_KEY_ORG1}'
```

Fig. 4. Example of CA private key

- Parameter hardcoded

Developers may leave some parameters hardcoded in configuration files especially for the development environment. The hardcoded parameter will cause difficulties in debugging and the experiments.

Fig. 4 shows an example of unreasonable hardcoded configuration and the reasonable one in a CA server node container. In listing 3, once the private key is changed, developer must modify the private key manually to fetch the new materials. In listing 4, developers can set any file name through the parameter without modifying any information manually.

- Parameter/component missing

The parameters or components missing will cause limited functions or failure. Developers may forget to configure some of parameters.

- *Yaml* syntax

*Yaml* file format is the official format provided in official documents of Fabric. All the *Yaml* configuration files must meet the *Yaml* syntax and principles.

- Docker / Kubernetes file syntax

Many developers choose Docker or Kubernetes for deploying Fabric network. Docker and Kubernetes use *Yaml* format file for startup containers. These configuration files must not only satisfy the *Yaml* syntax, but also the Docker and Kubernetes rules.

## B. Performance

Some configuration parameters are related with performance of the whole network. Unreasonable setting will cause low efficiency.

- BlockTime / BlockSize

BlockTime is the amount of time to wait before creating a block. BlockSize is the number of messages batched into a block. They are the most important parameters related to network performance.

If BlockTime is set too big, clients must wait a long time for every transaction in low pressure situation, although enough transactions can be batched into a block in high concurrency situation. Meanwhile if BlockTime is set too small, the situation goes to the contrary. Clients may not need to wait too long in the low pressure situation, but the ledger will be divided into more blocks in high concurrency situation. Otherwise, it will cause low efficiency especially for the network transmission.

In the situations with big BlockSize and low pressure, clients must wait for messages to reach the BlockSize to batched into a block, or the time is up to the BlockTime. But for the high concurrency situation, big BlockSize will make more messages into a block, that will reduce the network cost to improve efficiency of network. As a contrast, for the situation with small BlockSize and low pressure, clients will wait for less messages to reach BlockSize, that's much easier and more quickly. For the high concurrency, just like the small BlockTime situation, the block will be divided into many smaller blocks, it's very time-consuming for network transmission.

We think the effect on performance of BlockTime and BlockSize is also related with network load and network latency. However, too small or too big value should be unreasonable.

- Complex chaincode endorsement policy

Endorsement policies defines the peers in specific organizations which must endorse the execution of a transaction proposal. Complex endorsement policy will cause low efficiency, because clients must collect all endorse results from multiple peers to satisfy specific endorse policy, and it will take more time to commit transactions for validating every endorsement results. Complex endorsement policy transactions consume more computing resources, that will lead to low efficiency.

## C. Security

There are also some parameters which related with security of the whole network.

- Simple chaincode endorsement policy

Although complex chaincode endorsement policy causes low efficiency, simple chaincode endorsement policy will bring security problems. For example, when the policy is defined with '*OR(ORG1, ORG2)*', users could choose any single peer from either *ORG1* or *ORG2*. It will take users more cost for ensuring reliability and confidence of their target peer. Because the evil peer may tamper the endorsement result set and there is no other peer which can stop it. Otherwise, if more than one peer (especially from different organizations) is required for endorsement, it is safer because the adversary must invade all related peers.

- TLS on/off

Fabric supports Transport Layer Security (TLS) [8] for secure communication between entities, such as nodes and clients. Developers can choose whether TLS is turned on between each entity.

TLS has been introduced by Netscape in 1996. It's a kind of cryptographic protocol which is designed to provide transportation security in a computer network. Relied on the symmetric cryptography, no one can eavesdrop or tamper the messages between server and client.

In Fabric, user can turn on TLS for peer nodes, orderer nodes and peer CLI. Meanwhile, the related clients also must turn on TLS as well as the nodes which they communicate with.

By default[11], TLS client authentication is turned off both in the peer node and orderer node even when TLS is enabled. That means by default the node will not verify the certificate of a client, for example another node, application, or the CLI, during TLS handshake. So from a secure perspective point of view, developers should turn on both TLS and TLS client authentication.

- State database security

CouchDB is implemented as a separate database process in the outside of peer. Taking the Docker as the example, the peer container communicates with corresponding CouchDB container remotely. Generally developers configure the authentication information respectively for the peer container and CouchDB container, and make sure that they are consistent with each other.

In the official network samples, user name and password for the CouchDB container are all left empty. Although the original intentions may be that could facilitate developing and debugging, it's easy to miss these authentication information in production environment. And that will be very insecure because everyone can access the state database, even modify the chaincode data value to cause the inconsistent of ledger data.

- Consensus mechanism

As we know, Fabric supports multiple consensus methods, Solo, Kafka and Raft. Solo implementation is intended for test and only supports single orderer node. Kafka and Raft are crash fault tolerant(CFT) ordering service. However, there can be only single orderer node in every consensus method. In this situation the whole ordering service will be insecure because it can't be CFT. Once the single orderer node is malfunctioned or invaded, the whole network will be broken.

Developers should be advised not to use single ordering node in their network, whichever consensus mechanism was chosen.

V. SOLUTION FOR REASONABLE CHECK

We found that there is no related works or tools aiming at reasonableness check, especially for Fabric currently. So according to our knowledge and discovery about reasonableness problem in Fabric, we design a tool to check whether there is any unreasonableness in Fabric network configuration.

*A. Design*

Using *Yaml*, shell scripts and Docker is the most frequently used manner to configure and bootstrap a Fabric network, especially for the new developers. So our target is to analyze the network configuration files which are based on *Yaml* and shell scripts.

Based on the reasonableness problems descripted in Section IV, we design and implement 12 reasonableness check patterns to detect whether there are unreasonable state in Fabric configuration. The patterns are rule based to check 5 configuration files of Fabric network, which is mentioned before. It should be noted that for now our solution focuses on the static configuration of Fabric, and aims to help developers and administrators to optimize the network before it is all started.

The one reason we focus on the static configuration is that once network is running, many configurations will be immutable or very difficult to be modified. Furthermore, our solution can help people to understand their network more deeply, especially for the beginners and new developers of blockchain system.

Each pattern is corresponded to a specific reasonableness problem descripted in Section IV. The pattern contains a set of rules based on text matching and regular expression. We can detect and locate reasonableness problems by applying all the rules.

*B. Implementation*

Here is the steps to check reasonableness problems through patterns in our tools.

*1) Parse configuration files*

First, our tool collects and parse the 5 configuration files into pre-defined configuration items. The configuration files are *crypto-config.yaml*, *configtx.yaml*, *docker-compose.yaml*, *start.sh* and *scripts.sh*. According the usages of these 5 configuration files, the tool dumps and parses them as the network configurations together.

The dumped configurations will be stored in our additional database and waiting for analysis.

*2) Reasonableness check based on patterns*

We implement series of rules related with the 12 patterns. The rules are all based on the conditional statement, string matching and regular expression.

*3) Reasonableness check reports*

Once a rule is matched, a reasonableness tag will be generated in the check reports. Finally the check reports contain all the match result of every rules in every pattern. Also we provide a brief introduction and suggestion for this reasonableness problem.

Fig.5 is one instance of pattern result in the reports. The result contains the problem detail information, location, recommendation and level. There are 3 levels defined in the rules of every pattern, *Info*, *Warning* and *Error*. The importance is increased from *Info* to *Error*.

Fig.5. Report snippet of our tool

TABLE II. SAMPLE CONFIGURATION

| Project address | Name | Configuration file count |
|---|---|---|
| /yeasy/docker-compose-files/tree/master/hyperledger_fabric | Yeasy's fabric network sample | 43 |
| /hyperledger/caliper/tree/master/packages/caliper-samples/network | caliper samples | 50 |
| /IBM/build-blockchain-insurance-app | IBM/build-blockchain-insurance-app | 1 |
| /hyperledger/fabric-samples | fabric samples | 2 |
| /skcript/hlf-docker-swarm/tree/master/network | skcript/hlf-docker-swarm | 1 |
| /skcript/hyperledger-fabric-composer-multiorg-sample | skcript/hyperledger-fabric-composer-multiorg-sample | 1 |
| /MindtreeLtd/balance-transfer-java | MindtreeLtd/balance-transfer-java | 1 |
| /brucezhu512/blockchain-samples/tree/master/swarm | brucezhu512/blockchain-samples | 1 |
| /hyperledger-labs/fabric-multi-channel-network-samples.git | hyperledger-labs/fabric-multi-channel-network-samples | 1 |
| /nmatsui/fabric-payment-sample-docker.git | nmatsui/fabric-payment-sample-docker | 2 |
| /guoger/fabric-deployment.git | guoger/fabric-deployment | 1 |
| /hyperledger/composer/tree/166ae5cc365d8d524750a252e6c58a5094355167/packages/composer-tests-functional/hlfv1 | Fabric composer test sample | 1 |
| | Total count | 108 |

## VI. EXPERIMENTS

In this section, we perform a experiments to show that our proposed tool is very useful for Fabric network configuration.

### A. Experiment data collection

108 sample network configurations has been collected from the Github. We use the keyword 'Hyperledger Fabric' to query Fabric based projects manually. Table 2 shows the list of projects that contain one or more Fabric network configuration files. Some of these projects are the example network for learning Fabric, or samples in Fabric related tools such as Fabric composer. The others are sample networks of developers' use cases. The project addresses are all prefixed with *https://github.com*.

### B. Patterns execution and result analysis

Then, we execute all the rules in the patterns against the sample network configurations.

Table III shows the overall results of every pattern.

Overall, there are totally 504 reasonableness problems detected in these 108 Fabric networks. For detail, the most frequently problem is that TLS is turned off in the sample networks. We believe that is because most of networks are not from real project, they are just sample networks for study and development. The data quantity about endorsement policy is small. The reason is that the endorsement policy information is defined in *scripts.sh*, but most of the developers have not provided their *scripts.sh* file.

Some of problems will cause fatal error that can prevent the network from starting. *Yaml* syntax, Docker compose file syntax and component missing are these kinds of reasonableness problems. The results indicate that some developers publish their projects without audit and trial.

The tool gives *Info* when it detect the Solo consensus mechanism. There are 26 sample networks that are using Solo as the ordering service. The Solo implementation has been deprecated and may be removed in a future release of Fabric. So developers must pay attention to Solo.

53 sample networks choose LevelDB as the state database. Fabric provides great flexibility in the types of state database, but developers must consider which database should be chosen to satisfy their own requirements. Because state database cannot be modified once the network is running.

TABLE III. PATTERN RESULTS

| Patterns | Result count |
|---|---|
| State database choice | 53 |
| Inconsistent parameters | 0 |
| Parameter hardcoded | 7 |
| Component missing | 3 |
| Yaml syntax | 24 |
| Docker compose file syntax | 12 |
| BlockTime / BlockSize | 21 |
| Complex chaincode endorsement policy | 2 |
| Simple chaincode endorsement policy | 5 |
| TLS on/off | 343 |
| State database security | 8 |
| Consensus mechanism | 26 |

## VII. CONCLUSION

In this paper, we focus on the reasonableness problems of Hyperledger Fabric framework. As the permissioned blockchain system, there are various types of components in Fabric. The network preparation and configuration are also very complicated. Developers must deal with hundreds of parameters to configure and bootstrap their customized network. It's difficult and time consuming to make sure that all the network configurations are reasonable and satisfied application requirements, even for the veterans who are familiar with Fabric. Probably there are some unreasonable configuration parameters that may cause unexpected bad consequences. We define this kind of configurations as reasonableness problems.

We first discuss and summarize the potential reasonableness problems of Hyperledger Fabric according to different categories, functionality, security and performance. In consideration of the most common approach to build Fabric network, we think there are at least 12 reasonableness problems that may be caused in network configuration files. Then we implemented a check tool for reasonableness problems with 12 corresponding patterns. Every pattern contains a set of rules to match related problems. Finally we collect 108 sample network configurations from Internet, and run our tool on these sample networks as the experiment. The result of experiment shows that our tool is useful for checking and locating the reasonableness problems.

However, there is still many work to do about reasonableness check. 1) For now we only focus the static configuration before network running. In future, we will consider to focus on the dynamic network reasonableness check. 2) The experiment data are all just sample or experimental network configurations. We should collect more network configurations from real use cases. 3) All the 12 patterns is dedicated to Hyperledger Fabric, we think there are more common patterns that are appropriate for other permissioned blockchain system.